# CONSTRUCTION OF MICROACCELERATIONS FRACTAL MODEL ON THE BASIS OF WEIERSTRASS-MANDELBROT FUNCTION

**Sedelnikov A.V., Podlesnova D.P.**
(Samara state aerospace university)
axe_backdraft@inbox.ru

In work the opportunity of construction fractal estimation of low-frequency component microaccelerations with the help of the valid part to Weierstrass-Mandelbrot function at identically zero phase is statistically proved. The proof will be carried out with the help of correlation factor and nonparametric rang criterion Cox - Stuart.

The microaccelerations problem is an actual problem of space materiology on an extent more than 30 years already. After unsuccessful attempt of creation technological spacecraft "NIKA-T" where the microaccelerations low-frequency components module should not exceed $2 \cdot 10^{-6} g$, designing and release of a space mini-factory on base spacecraft "OKA-T" with a level of low-frequency components is planned is not higher $10^{-7} g$.

In this connection also other technological projects there is a task of an estimation of the microaccelerations module on technological spacecrafts. Such task, probably, is insoluble in the common statement as on two structurally identical spacecraft the microaccelerations level will be various because of the start conditions of these devices is differenced. Influence of the space environment on spacecrafts will be various also during orbital flight (solar pressure, micrometeorites etc.) is especial when the question is about $10^{-7} g$. Therefore it is meaningful to present an microaccelerations estimation incorporated in itself spacecrafts design-layout circuit (DLC). It just that microaccelerations part which can be operated at technological spacecraft creation at a stage of their designing.

This work is a part of a works cycle on creation fractal models of microaccelerations low-frequency components incorporated in spacecraft DLC. It is supposed that spacecraft has the big elastic elements (panels of solar batteries, a radiator etc.). Dynamics of change of the microaccelerations module after individual operation of managing rocket engines of orientation system spacecraft (MRE) is investigated. In this case the microaccelerations field inside spacecraft is made by fluctuations of spacecraft elastic elements.

Average value of microaccelerations low-frequency components which within the framework of a considered task depend on MRE moment and accelerative-mass characteristics of spacecraft elastic elements is offered to model with the help of the valid part of Weierstrass-Mandelbrot function at identically zero phase (WMF):

$$\mathrm{Re}\, W(t) = C(t) = \sum_{n=-\infty}^{n=+\infty} \frac{1-\cos\left(b^n t\right)}{b^{(2-D)n}} \qquad (1)$$

The purpose of work the statistical proof of suitability of function (1) for an estimation of average value microaccelerations low-frequency components. For this purpose we shall consider an estimation of correlation between change of average value microaccelerations low-frequency components at change of moment MRE and accelerative-mass characteristics of spacecraft elastic elements and change of WMF average value at change of parameters of function, preliminary having made a number of remarks.

The opening remark is connected by that WMF valid selfafinity the range of parameter change $t$ function has no special value. In work [1] it is offered to identify it with dimensionless time, and value $t = 0$ to count the beginning of technological process and $t = 1$ its ending.

The second remark concerns that microaccelerations low-frequency components is weak dampped. Researches [3] show that in time appropriate to realization of technological process average value and average disorder around of average value of this component can be counted constants. In this sense change of the microaccelerations module in time is adequate to concept of a random variable. In this connection analysis WMF was carried out and restrictions on parameters of function (fractal dimension $1{,}99 \le D < 2$ for 5 %-s' significance values) are imposed so that it also corresponded to concept of a random variable of sense of a dispersion and average value constancy.

The analysis was carried out under the following circuit. It is known that WMF is growing function [4]. Therefore in any sample its edges should differ most strongly. Two matched selection were made of sample in 1000 points on 100 points everyone. Then the hypothesis of uniformity of these samples which was checked on 5 %-s' significance value with the help of Smirnov`s criterion (conformity of distribution laws), Fisher`s criterion (equality of selective dispersions) and Wilcoxon-Mann-Whitney`s (equality of average values) was put forward. Results of statistical check WMF for $b = 0{,}7$ and it various values are given fractal dimension on fig. 1-3. The figure in square brackets on the diagram means number nines after a comma (WMF fractal dimension should be strictly less than two).

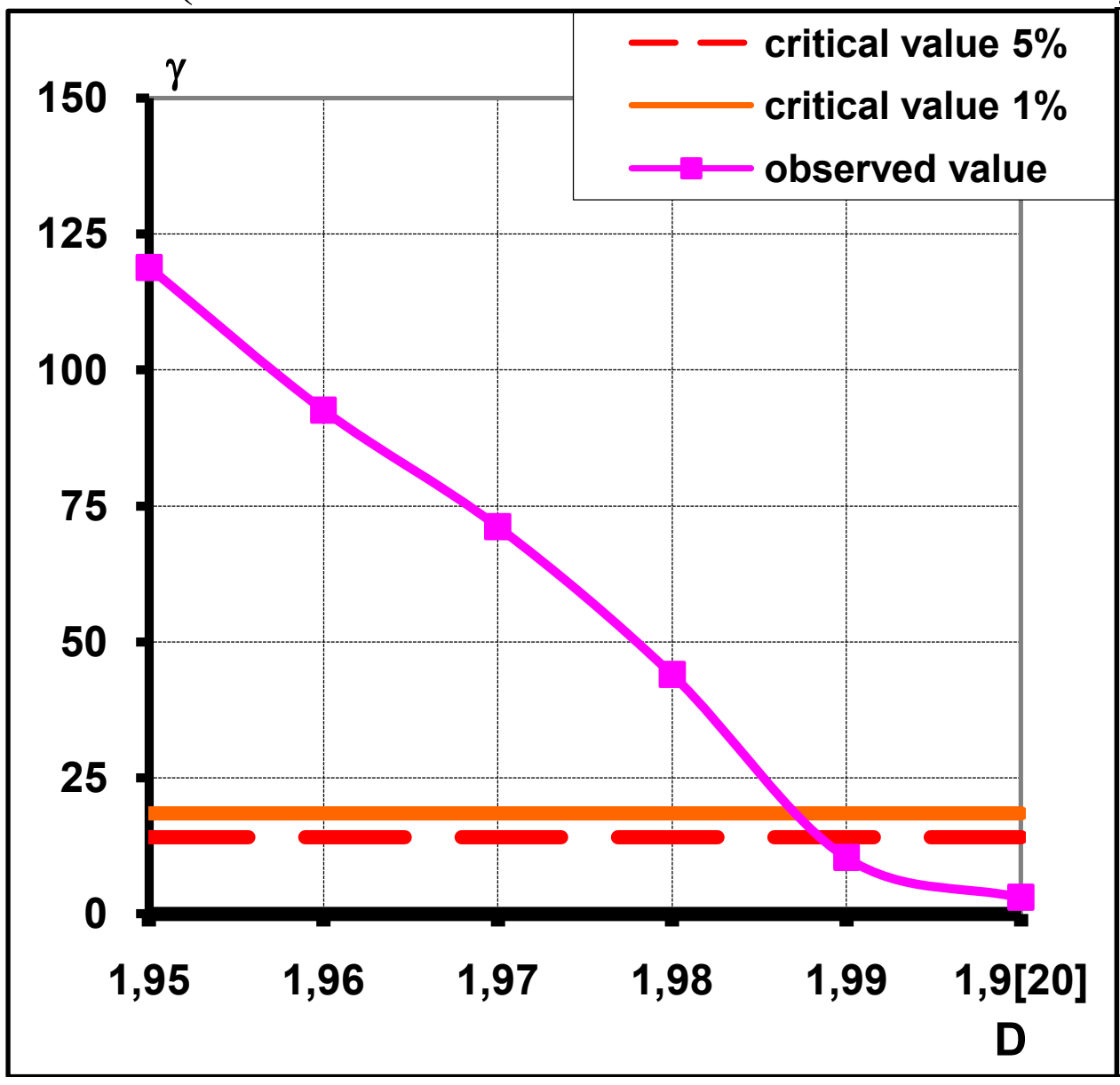

Fig. 1 - Dynamics of Smirnov`s criterion observably values change from fractal dimension

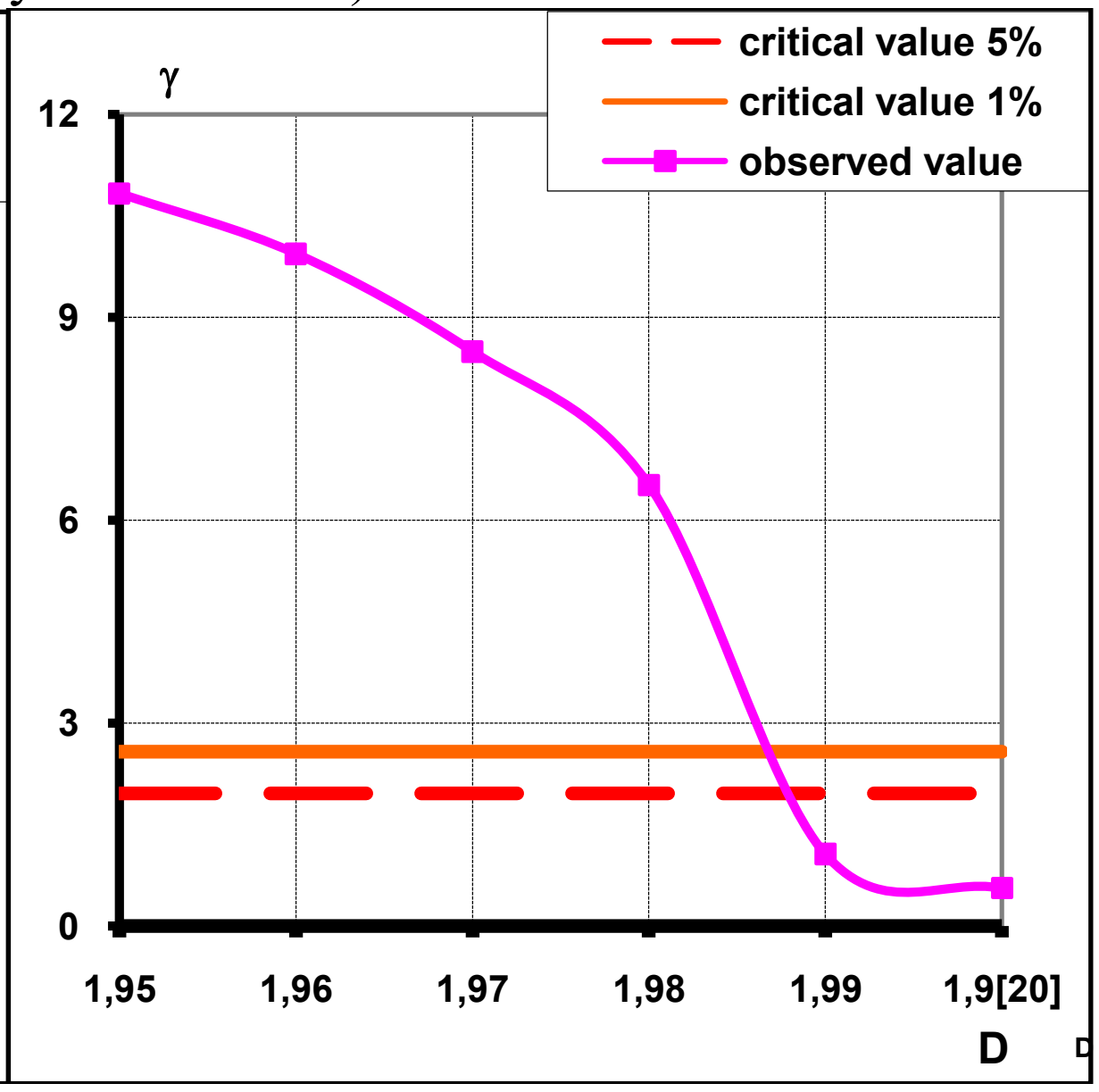

Fig. 2 - Dynamics of Wilcoxon-Mann-Whitney`s criterion observably values change from fractal dimension

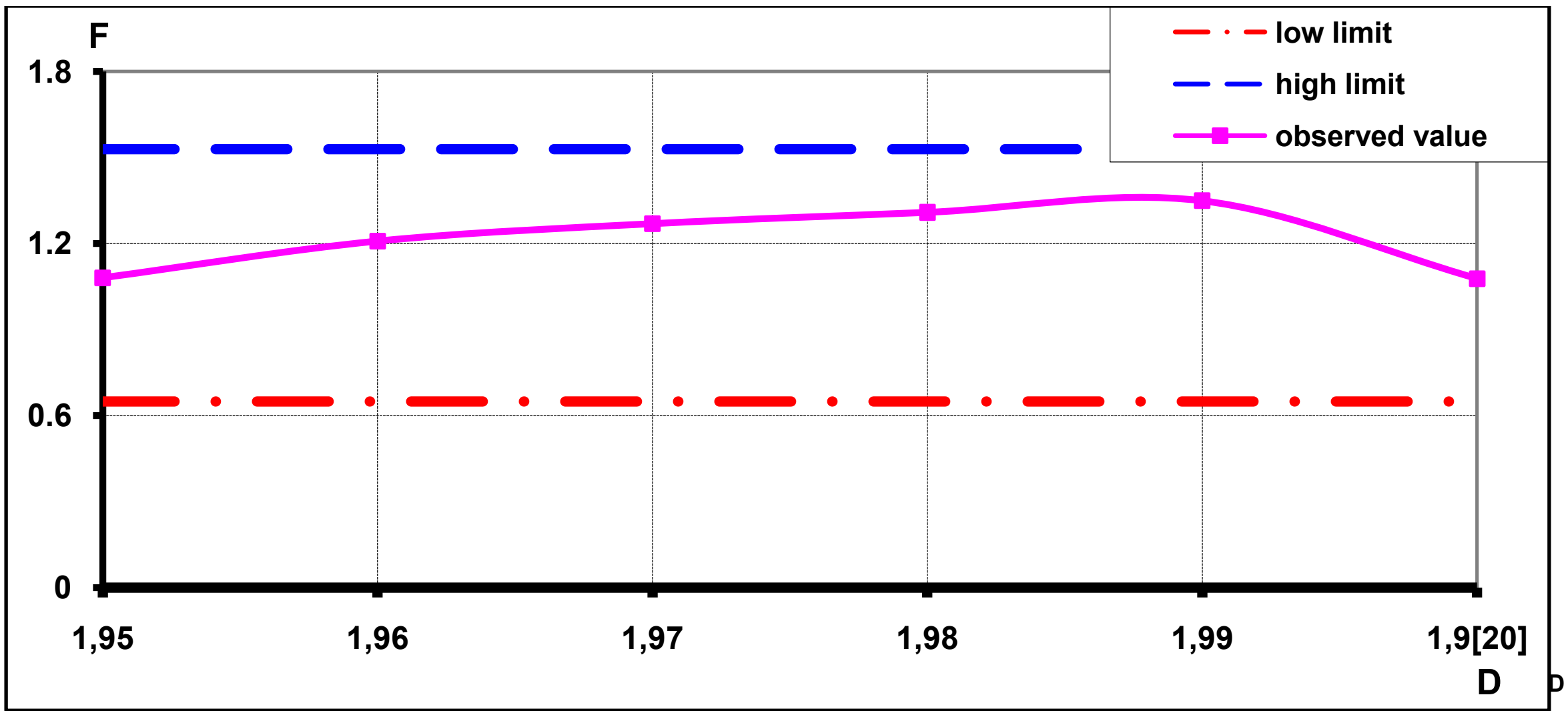

Fig. 3 - Dynamics of Fisher`s criterion observably values change from fractal dimension

Similar results are received and for other values of WMF scale parameter *b*.

The Third remark is connected to reception of microaccelerations values. For reception of microaccelerations average value spatial rotation of spacecrafts such as "NIKA-T" having three elastic elements (model [5]) was considered. In model [ 5 ] elastic elements are submitted by spatial plates. It is taken into account first six forms of fluctuations.

The Estimation of correlation was carried out with the help of correlation factor:

$$r = \frac{\sum_{i=1}^{n}\left[\left(x_i - \bar{x}\right)\left(y_i - \bar{y}\right)\right]}{\sqrt{\sum_{i=1}^{n}\left(x_i - \bar{x}\right)^2 \sum\left(y_i - \bar{y}\right)^2}}. \quad (2)$$

In ( 2 ) values $x_i$ are understood as microaccelerations average values received on model [ 5 ] and as values $y_i$ WMF average values. Dependences of microaccelerations average value from MRE moment for various values of the spacecraft accelerative-mass characteristics received on model [ 5 ] are given on fig. 4-6.

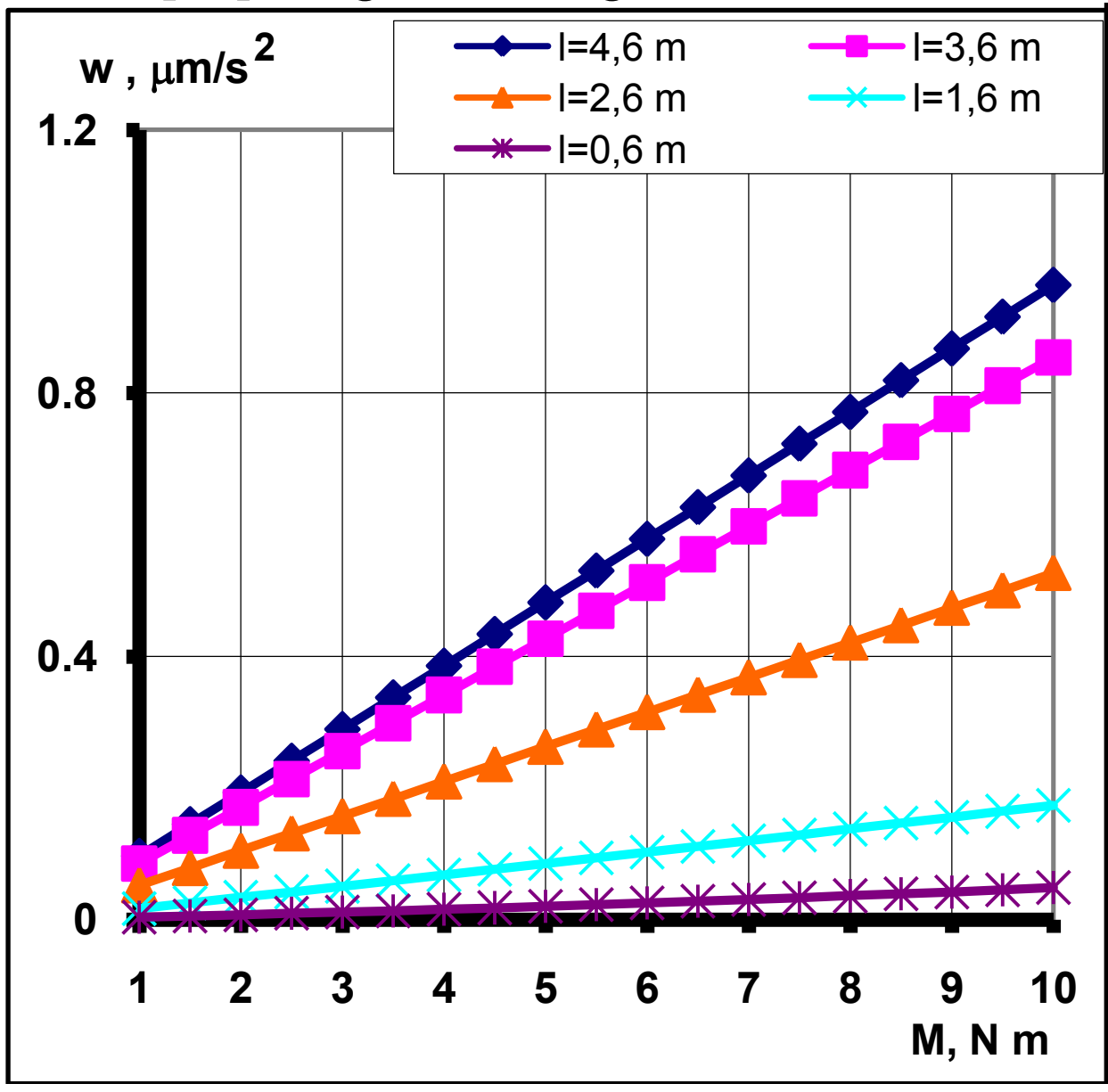

Fig. 4 - Dependence of average value of microaccelerations low-frequency components from MRE moment at spacecraft mass 6500 kg and running weight elastic elements 20 kg / m

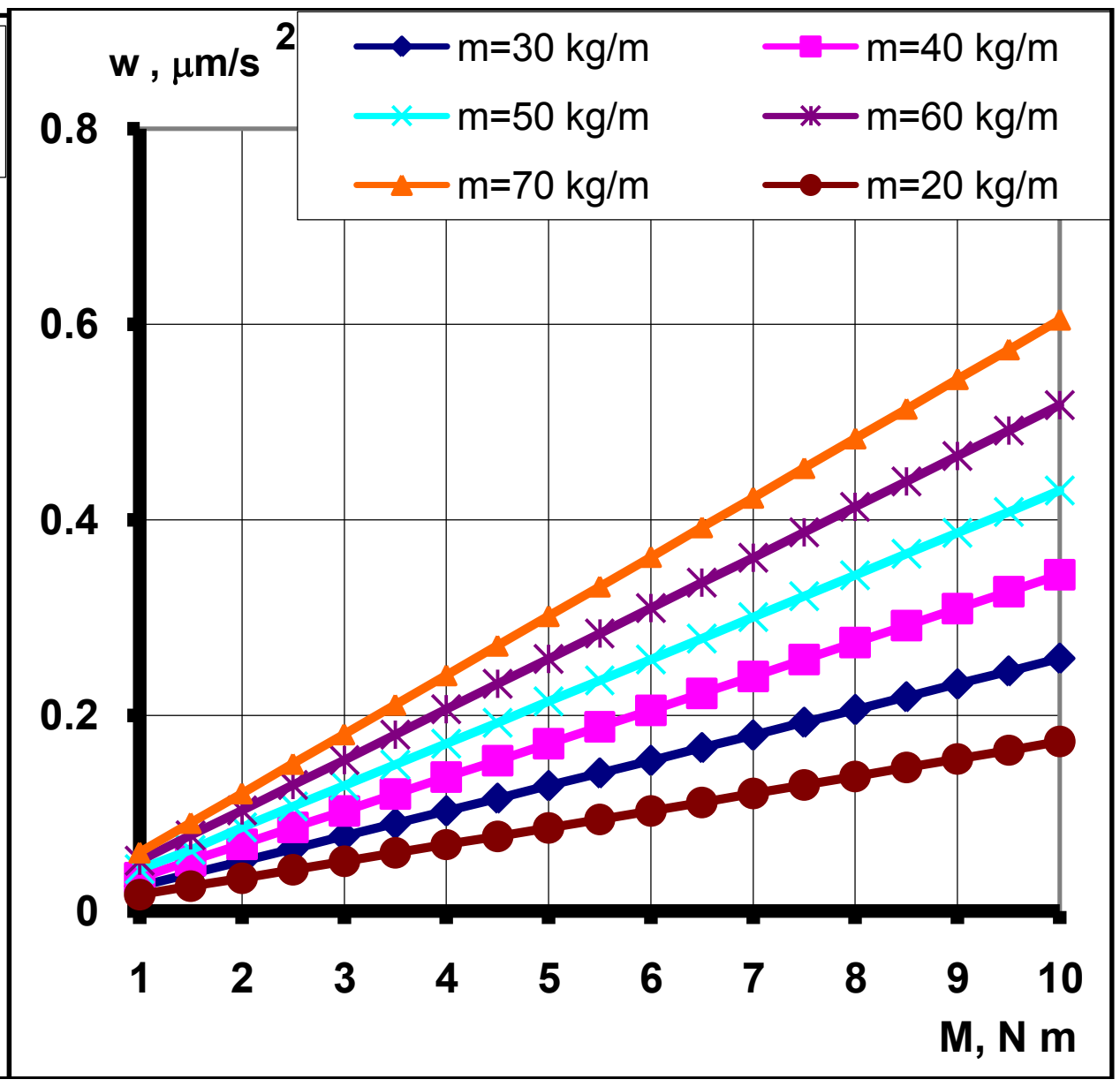

Fig. 5 - Dependence of average value of microaccelerations low-frequency components from MRE moment at spacecraft mass 6500 kg and length elastic elements 1,6 m

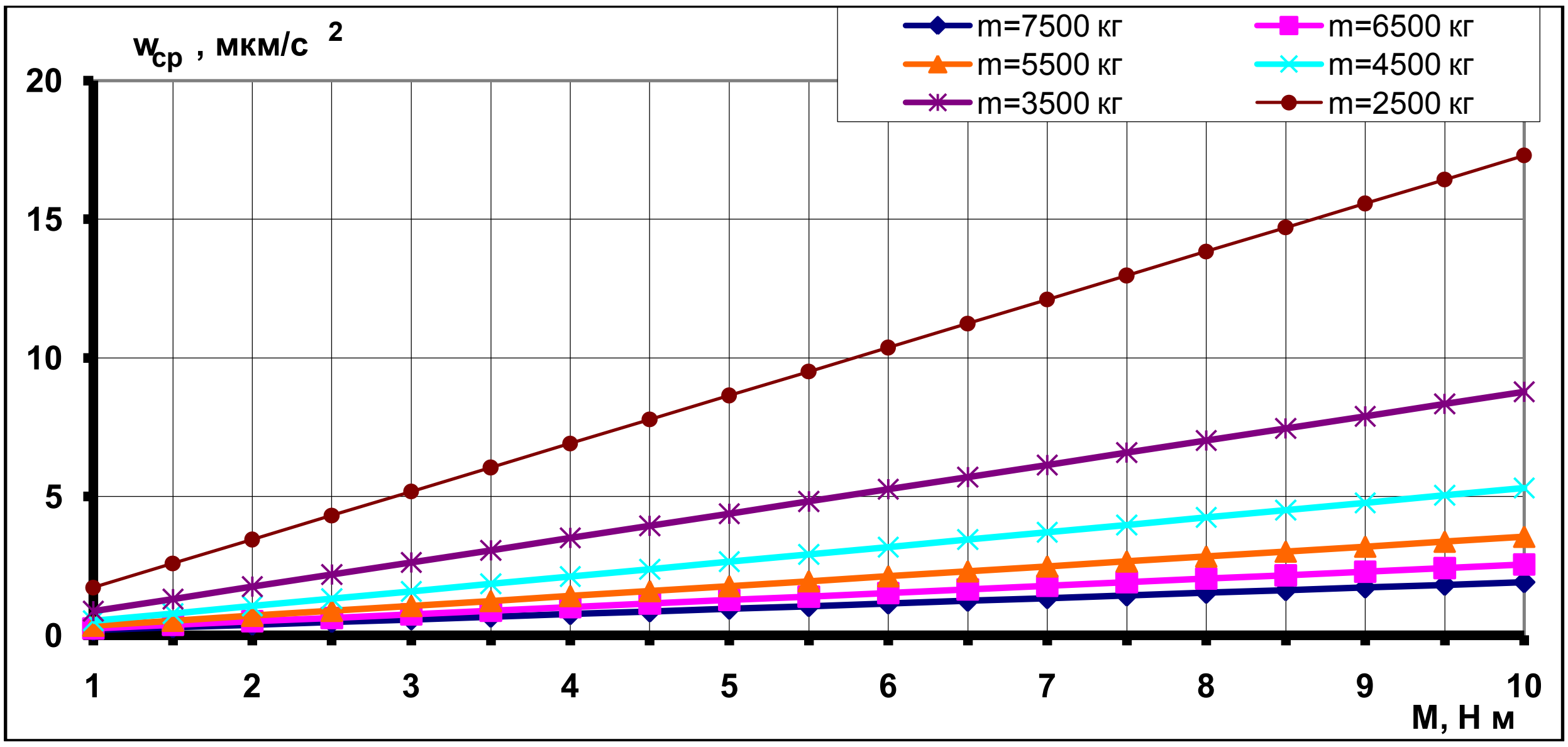

Fig. 6 - Dependence of average value of microaccelerations low-frequency components from MRE moment at next elastic elements parameters running weight 48,9 kg/m and length 4,6 m

Dependence of WMF average value on its parameters is submitted on fig. 7.

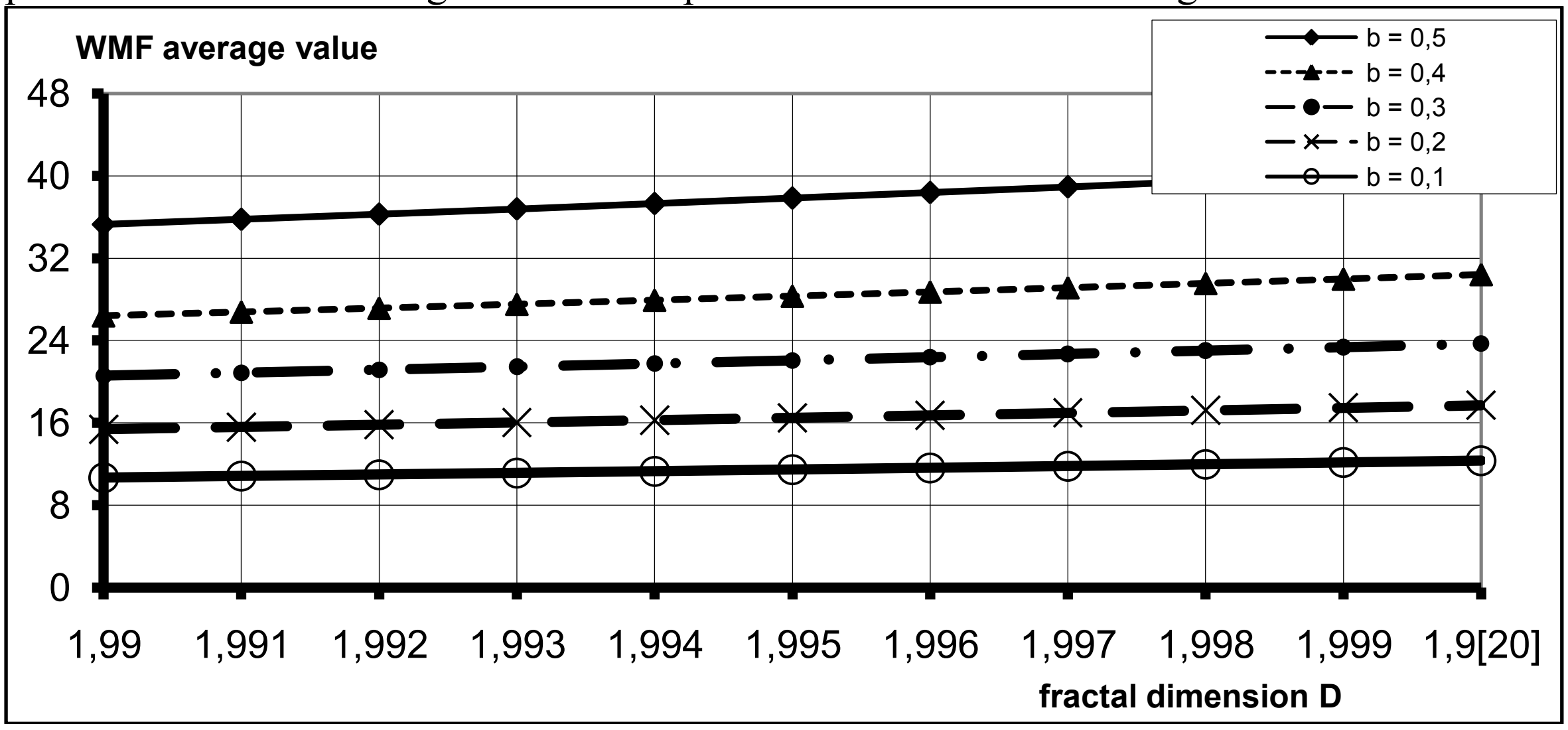

Fig. 7 - Dependence of WMF average value on it fractal dimensions at various values of scale parameter

The analysis shows that values of correlation factor (2) is in limits from 0,80 up to 0,90 for various samples. It confidently allows to accept a hypothesis about linear correlation between microaccelerations average value and WMF average value on 5 %-s' significance value (critical value 0,537).

However the correlation factor is unstable to displacement from the normal law of distribution. Therefore for a final conclusion about applicability WMF for an estimation of microaccelerations average value Cox-Stuart`s [6] is applicable nonparametric rang criterion. We shall consider correlation of the dependences submitted on fig. 4 and 7. We shall make samples on 55 points and we range them on WMF average value (tab. 1).

Tab.1

| **$x_i$** | **$y_i$** | **$x_i$** | **$y_i$** | **$x_i$** | **$y_i$** | **$x_i$** | **$y_i$** | **$x_i$** | **$y_i$** | **$x_i$** | **$y_i$** |
|---|---|---|---|---|---|---|---|---|---|---|---|
| 0,004 | 10,67 | 0,006 | 10,83 | 0,008 | 10,98 | 0,010 | 11,14 | 0,011 | 11,30 | 0,014 | 11,46 |
| +1 | | +1 | | +1 | | +1 | | +1 | | +1 | |
| 0,016 | 11,63 | 0,018 | 11,80 | 0,020 | 11,98 | 0,023 | 12,16 | 0,025 | 12,34 | 0,017 | 15,37 |
| +1 | | +1 | | +1 | | +1 | | +1 | | +1 | |
| 0,025 | 15,58 | 0,034 | 15,80 | 0,043 | 16,02 | 0,051 | 16,25 | 0,060 | 16,48 | 0,068 | 16,72 |
| +1 | | +1 | | +1 | | +1 | | +1 | | | |
| 0,077 | 16,96 | 0,085 | 17,20 | 0,094 | 17,45 | 0,103 | 17,70 | 0,053 | 20,57 | 0,079 | 20,86 |
| 0,105 | 21,15 | 0,131 | 21,45 | 0,158 | 21,75 | 0,184 | 22,06 | 0,210 | 22,38 | 0,237 | 22,70 |
| 0,263 | 23,02 | 0,289 | 23,36 | 0,316 | 23,70 | 0,085 | 26,41 | 0,128 | 26,78 | 0,171 | 27,15 |
| 0,214 | 27,53 | 0,256 | 27,92 | 0,299 | 28,32 | 0,342 | 28,72 | 0,384 | 29,14 | 0,427 | 29,55 |
| 0,470 | 29,98 | 0,512 | 30,42 | 0,096 | 35,25 | 0,145 | 35,75 | 0,193 | 36,25 | 0,241 | 36,77 |
| 0,289 | 37,29 | 0,337 | 37,82 | 0,386 | 38,37 | 0,434 | 38,92 | 0,482 | 39,48 | 0,530 | 40,06 |
| 0,579 | 40,65 | | | | | | | | | | |

In tab. 1 designations the same, as in expression for classical correlation factor. We shall break the given sample on three matched selection: 17; 21; 17. Also we shall analyse rang correlation having compared two extreme matched selection. Apparently from tab. 1, all values $x_i$ from the first matched selection less similar from the third. Therefore the rang statistics value will look like:

$$T = +1 + 1 + 1 + 1 + 1 + 1 + 1 + 1 + 1 + 1 + 1 + 1 + 1 + 1 + 1 + 1 + 1 = +17$$

The critical statistics value for 5 %-s' significance is equal 9 [ 6 ]. Therefore the hypothesis about correlation is accepted. Moreover, value $T$ is maximum for the given sample. It testifies to presence of strong linear correlation between microaccelerations average values and WMF average values. Similar results can be received considering other samples and investigating correlation between dependences fig. 5 and fig. 7, and also fig. 6 and fig. 7.

Thus, with the help of correlation factor and Cox–Stuart`s criterion the opportunity of modelling of microaccelerations low-frequency components average value with help WMF in the statement suggested is proved.

[1] *Sedelnikov A.V.* Qualitative identification of Weierstrass-Mandelbrot function parameters at an microaccelerations estimation. Science in the higher school: problems of integration and innovations. Materials of VII International scientific conference. Moscow. 2007. 42-52.

[2] *Sedelnikov A.V., Podlesnova D.P.* Main principles of construction of analytical dependence of Weierstrass-Mandelbrot function parameters for an microaccelerations estimation. Successes of modern natural sciences. 2006. **12**, 82-83.

[3] *Sedelnikov A.V., Byazina A.V., Ivanova S.A.* Microaccelerations statistical researches at presence weak decrement fluctuations of spacecraft elastic elements. Scientific readings in the Samara branch of the Russian academy of education. **part 1**, Natural sciences. M.: university of RAE. 2003. 137–158.

[4] *Berry M.V., Lewis Z.V.* On the Weierstrass-Mandelbrot fractal function. Proc. R. Soc. **A370**, London. 1980. 459-484.

[5] *Avramenko A.A., Sedelnikov A.V.* Modelling of a field of residual microgravitation onboard orbital spacecraft. News of high schools. Aviation engineering. 1996. **4**, 22-25.

[6] *Kobzar A.I.* Applied mathematical statistics. M: Fyzmathlit. 2006.

Address for connection: Russia, state Samara, 443026, p.b. 1253, Dr. Andry Sedelnikov
Address on Russian: Россия, г. Самара, а/я 1253.
E\mail: axe_backdraft@inbox.ru